\documentclass[conference]{IEEEtran}

\usepackage[T1]{fontenc}
\usepackage{amsmath}
\usepackage[cmintegrals]{newtxmath}

\usepackage{float}
\usepackage{graphicx}
\usepackage[hidelinks, bookmarks=false]{hyperref}
\usepackage{subcaption}
\usepackage{tikz}
\usepackage{times}
\usepackage{xcolor}

\usetikzlibrary{positioning}

\edef\restoreparindent{\parindent=\the\parindent\relax}
\usepackage{parskip}
\restoreparindent
\title{\huge Deployment of a Blockchain-Based Self-Sovereign Identity}
\author{\IEEEauthorblockN{Quinten Stokkink}
\IEEEauthorblockA{Distributed Systems\\
Delft University of Technology\\
Email: q.a.stokkink@tudelft.nl
}\and\IEEEauthorblockN{Johan Pouwelse}
\IEEEauthorblockA{Distributed Systems\\
Delft University of Technology\\
Email: j.a.pouwelse@tudelft.nl
}}

\begin{document}
\maketitle

\begin{abstract}
Digital identity is unsolved: after many years of research there is still no trusted communication over the Internet.
To provide identity within the context of mutual distrust, this paper presents a blockchain-based digital identity solution.
Without depending upon a single trusted third party, the proposed solution achieves passport-level legally valid identity.
This solution for making identities Self-Sovereign, builds on a generic provable claim model for which attestations of truth from third parties need to be collected.
The claim model is then shown to be both blockchain structure and proof method agnostic.
Four different implementations in support of these two claim model properties are shown to offer sub-second performance for claim creation and claim verification.
Through the properties of Self-Sovereign Identity, legally valid status and acceptable performance, our solution is considered to be fit for adoption by the general public.
\end{abstract}

\begin{IEEEkeywords}
Blockchain, Distributed Ledger, Digital Identity, Zero-Knowledge Proof.
\end{IEEEkeywords}

\section{Introduction}
Through businesses and institutions existing in an ecosystem of mutual distrust, identity has gotten fragmented.
This has led to large amounts of our identity data being duplicated onto the servers of authorities.
Current identity solutions require us to hold many plastic cards which we use to identify ourselves or pay with.
We need to have a username and password for many different websites.
This means that the individuals whose data is stored have no -or little- control over this data.
Furthermore, these servers which collect privacy sensistive data become prime targets for attacks: honeypots.

We still have a digital identity problem, even though solutions have already been proposed for this identity problem many decades ago.
The concept of public keys has been around since 1976, when Diffie and Hellman created their (Diffie-Hellman) key exchange method.
Based on this public key concept, Pretty Good Privacy (PGP) has been around since $1991$~\cite{zimmermann1991wrote}.
However, it has also been known for 19 years~\cite{whitten1999johnny} that PGP, made 27 years ago, is not being adopted.
To this day this adoption problem persists.
There has yet to emerge a digital identity solution that solves the digital identity schizophrenia.

Public key technology forms the birth of both the problem and solution for digital identities.
This is all of the cryptography we require to securely communicate with other identities.
After all, public keys gave us confidentiality and integrity of data.
There is just the issue of having to tie such a public key to an identity: a need for a Public Key Infrastructure.
This is the real issue we have tried to solve for all of these years.
Nobody trusts the Public Key Infrastructure of his competitor.
As such, a model which relies on central servers is doomed to fail.
This is why we do not see initiatives such as \textit{keybase.io} as a move forward to solve the identity problem.

More recent work has attempted to solve identity without a central trusted third party.
The concept is that one simply exists, therefore one has a digital identity.
The identity is then built, managed and used by the owner of the identity.
No longer is the identity owner beholden to a single identity authority.
This is called \textit{Self-Sovereign Identity}.
Now, the businesses and institutions no longer need to trust each other, they just need to trust the user.

One might be tempted to say that the concept of \textit{Self-Sovereign Identity} finally solves our 27 year old digital identity problem.
However, whereas the businesses and institutions no longer need to trust each other, they also do not trust the user to be truthful.
This paper will build upon the findings of previous \textit{Self-Sovereign Identity} initiatives such as \textit{Sovrin}\footnote{https://sovrin.org/} and \textit{uPort}\footnote{https://www.uport.me/} and leverages a blockchain to solve this trust problem.
By utilizing a blockchain, users can be tied to the claims they make and hereby be caught red-handed if they attempt to cheat the system.
Users can be caught committing identity fraud.

In contrast to the previously mentioned solutions, this paper will present an academically pure model for \textit{Self-Sovereign Identity}.
In our model there does not need to be a foundation assigning infrastructure.
In our model there is no possibility of the blockchain owner or $51\%$ of the network to hard-fork a user out of existence~\cite{atzei2017survey}.
There will be no third party in control of attributes: there will be no vendor lock-in.
A key property of our model is open enrollment: any user can simply start using the solution without asking for permission.

The model of this paper has been created in cooperation with the Dutch National Office for Identity Data (Ministry of the Interior and Kingdom Relations).
As such it will be the second digital identity model in the world to be sanctioned by a government (after Estonia).
Through this cooperation the solution of this paper can also leverage passport-level biometry. 
The key contribution of this paper is to provide these passport-level attributes for an academically-pure secure open \textit{Self-Sovereign Identity} standard.
This paper describes the concepts of a pilot which will be deployed to Dutch citizens in the latter half of 2018.

\section{Problem Description}
The first half of the problem we observe in the current identity ecosystem, is the fact that identity holders should also be the identity owners.
This first half can be more formally described as the need for \textit{Self-Sovereign Identity}.
The second half of the problem consists of the passport-level attributes in this identity.
In other words, identities which are recognized by governments and therefore have legal value.
In the context of blockchains we can formalize this second half of the problem as the need for legally valid signatures.
This section will first explain the problems of \textit{Self-Sovereign Identity} and then the section will follow up with the description of legally valid signatures.

\subsection{Self-Sovereign Identity}
The name \textit{Self-Sovereign Identity} already hints at the fact that in this model users have full control over their identity.
We will now explain what it means to be a sovereign of one's self, or in other words, to have full control over one's own identity.
Thankfully, in the essay \textit{The Path to Self-Sovereign Identity} from 2016~\cite{allen2016principles}, Christopher Allen has created a comprehensive list of properties of Self-Sovereign Identity.
We prefer this list of properties over other works (like Zyskind et al.~\cite{zyskind2015decentralizing}), as it is more extensive.
For the full argumentation we refer the reader to the essay; we will quote C. Allen's 10 properties of Self-Sovereign Identity for the reader's convenience:

\begin{enumerate}
\item \textbf{Existence.} \textit{Users must have an independent existence}.
\item \textbf{Control.} \textit{Users must control their identities}.
\item \textbf{Access.} \textit{Users must have access to their own data}.
\item \textbf{Transparency.} \textit{Systems and algorithms must be transparent}.
\item \textbf{Persistence.} \textit{Identities must be long-lived}.
\item \textbf{Portability.} \textit{Information and services about identity must be transportable}.
\item \textbf{Interoperability.} \textit{Identities should be as widely usable as possible}.
\item \textbf{Consent.} \textit{Users must agree to the use of their identity}.
\item \textbf{Minimalization.} \textit{Disclosure of claims must be minimized}.
\item \textbf{Protection.} \textit{The rights of users must be protected}.
\end{enumerate}

\noindent By leveraging a blockchain structure, one can observe that some of these properties are intrinsically fulfilled.
By running a consensus mechanism, a global truth can be established (according to at least $51\%$ of the network) for each of the published attributes.
This consensus mechanism therefore provides the \textbf{transparency} we were seeking.
Furthermore, due to the append-only nature of blockchains (once the information has been accepted after a consensus round) the blockchain will also provide the \textbf{persistence} of the published attributes.

By delving into more specialized blockchain structures we can trim down this list of requirements even further.
Chain structures like that of \textit{TrustChain}~\cite{otte2017trustchain} and \textit{The Tangle}~\cite{popov2016tangle} provide a single chain per single user/application, or in this case per identity.
Full \textbf{control} is then given in these structures to the user which owns one of these personalized chains, without limiting the derived properties of transparency and persistence from the general blockchain structure.
By allowing the identity to be based on a personalized blockchain, which a user can create without any permissioning, this gives the user the property of \textbf{existence}.
Also, because the user creates and maintains his own blockchain, this means the user has \textbf{access} to this data and has to synchronize his data with the rest of the world through his own effort.
Through this conscious effort of sharing data the user gives his explicit \textbf{consent}.

In conclusion, many of the properties of Self-Sovereign Identity are intrinsically provided by leveraging a personalized blockchain structure like TrustChain or The Tangle.
By leveraging such a personalized blockchain structure, we can determine the properties the \textit{claims} should fulfill to provide for all of the $10$ properties of Self-Sovereign Identity.
The list of open problems the claim format should solve is as follows:
\begin{itemize}
\item \textbf{Portability}
\item \textbf{Interoperability}
\item \textbf{Minimalization}
\item \textbf{Protection}
\end{itemize}

\noindent This paper will add one further requirement for a Self-Sovereign Identity system beyond these \textit{10 principles of Self-Sovereign Identity}.
Claims are not worth anything if they can not be shown to hold true.
These generic claims therefore also need to be \textbf{provable}.

\subsection{Legally valid signatures}
A strict requirement for the adoption of any identity system for government-level use, is the use of legally valid signatures.
That is to say that the signature algorithm and the corresponding keys to use this identity should be of sufficient quality.
Sufficient quality is a vague directive in itself, which we will interpret as usage of keys  sampled from elliptic curves which have been approved for government use by an institute for standards, like the National Institute of Standards and Technology.
Of course, this list of ``safe'' curves updates regularly and as such there are also many websites devoted to keeping track of this (for instance https://safecurves.cr.yp.to/).

Next to the choice of a ``safe'' key, users of the system should also be recognized by a government institution before their signatures become legally valid.
This means that the goverment should recognize the existence of the user of the ``safe'' key.
Here \textit{the government} is any other identity (or keypair if you will) which is associated with the government.
This association might be realized through a hierarchical construction or be sampled from a list of trusted keys.
In our model, we could also see this recognition as just another claim attested to by the government.

Alongside the protection of citizens, legal signatures are also required for \textit{auditing}.
That is to say that the global state of the network should be readable for the government to detect cheating.
When a blockchain is utilized as an identity building platform, double-spending attacks are now a criminal offense: identity fraud.
If a citizen commits identity fraud he should be punishable by law.
But, again, this can only happen if the citizen is legally recognized to be bound to his signature.

\section{Design}
\label{sec:design}
As stated before, half of the design for our true Self-Sovereign Identity model leans on the properties of personalized blockchain structures like TrustChain or The Tangle.
The other half will consist of claim structures which should be \textbf{portable}, \textbf{interoperable}, \textbf{minimized}, \textbf{protective} and \textbf{provable}.
We will first start by examining a claim model which satisfies these criteria, afterwards we will present the modeling of these claims into a blockchain structure.

The claims which will be presented do not hold any intrinsic truth.
Users can make false claims and users can erase their identity (also called a \textit{whitewashing} attack~\cite{feldman2006free}).
That is to say that information delivered by a user should only be regarded as the truth, if the user can provide proof for the claim.
Establishment of truth will have to involve multiple parties, which provide \textit{attestations}~\cite{azouvi2017secure}.
The trust in other parties attesting to the truth of a user's claim then adds up to the trust in the user's claim.

\subsection{Generic provable claims}
\label{sec:designgenclaims}
The most important property of Self-Sovereign Identity, regarding the design of claims, is that of \textbf{protection}.
The user's right to privacy and the right to be forgotten can be joined into the need for a mechanism to disclose information to specific parties on demand.
Furthermore, once shared, the information the other party receives should not be fit for re-use.
In other words, the received information should hold truth only for the party the information was disclosed to.
This lends itself nearly perfectly to a Zero-Knowledge Proof structure~\cite{goldreich1986proofs}.

Usage of Zero-Knowledge Proofs also helps in the area of \textbf{minimalization}.
By designing the Zero-Knowledge Proofs accordingly, one could limit the access to a single attribute to only be provably within a certain range~\cite{peng2010efficient}.
This is useful when -for instance- sharing minimum levels of income.
One could prove a certain level of income without having the public attribute stating the actual amount the income exceeds.
The existence of the attestation of ``income is higher than \$1'' leaks the fact that the income is higher than ``\$1''.

These Zero-Knowledge enabled claims then form the \textbf{portable} and \textbf{interoperable} building block we required.
The claims do not require any blockchain to be evaluated and may be shared with other platforms.
But, even though these claims would be able to be shared, they are not generic.
The claims are still missing a format for ease of use and easy upgrading: the claims are still missing metadata.
Our solution for the claim metadata is given in \autoref{tab:claimmetadata}.

\subsubsection{Name}
The claim name should be a globally recognized value for identification of a particular attribute.
To share infrastructure between attesting parties and verifying parties, a naming convention should be present.
In principle this convention, or standardization, need only happen between the verifying and the attesting party.
However, issues can also arise when one attesting party has a different perception of an attribute than another attesting party.
Should, for example, the attribute ``name'' be a full name or a family name?

\subsubsection{Timestamp}
The timestamp is a means for fraud detection through auditing.
By withholding revocations, users may attempt to commit identity fraud.
This can be mitigated by storing the signature of the identity owner over a timestamp $T_1$ and the tail (last hash) of the blockchain.
A verifying party can then prove that a revocation existed at a certain time $T_0 < T_1$ and hereby provide legal basis for identity fraud.
Furthermore, timestamps can also be used to issue unique challenges for proof of key ownership: authentication using claims.
Through this challenge, proof of ownership of a second key can then be shown by producing a signature using this second key.
In other words, this allows for deferred \textit{two-factor authentication} through claims.

\subsubsection{Validity term}
By attesting to the truth of values of attributes of others, the attesting parties stake their reputation.
If an attesting party provides an attestation for an attribute and it turns out to be wrong, one might even say that the attesting party is complicit in identity fraud.
As, in the real world, values of attributes tend to change over time we can leverage a validity term.
After this term has expired, any attestations provided for this claim are to be viewed as void in the eyes of the beholder.
The risk of using such an expired claim is then fully deferred to verifying party.

\begin{table}[!t]
\renewcommand{\arraystretch}{1.3}
\caption{Claim metadata}
\label{tab:claimmetadata}
\centering
\begin{tabular}{ p{.25\columnwidth} | p{.65\columnwidth} }
    \hline
    \textbf{Field} & \textbf{Description} \\  \hline
    Name & The name of the attribute\\
    Timestamp & The time of claim creation\\
    Validity term & The time after which the claim is no longer valid\\
    Proof format & The type of proof for the claim\\
    Proof link & The strong link to the proof for the claim\\
    \hline
\end{tabular}
\end{table}

\begin{figure*}[t]
\centering
\begin{subfigure}{0.45\textwidth}
\centering
\begin{tikzpicture}
\node[rectangle,draw,fill=lightgray!30!white,minimum width=.6\columnwidth](pchainentries){Previous Chain Entries};
\node[rectangle,draw,fill=white,minimum width=.6\columnwidth,below=.7cm of pchainentries](metadata){Identification Metadata};
\node[rectangle,draw,fill=white,minimum width=.6\columnwidth,below=0cm of metadata](attestorsign){\textit{Signature of Attesting Party}};
\node[rectangle,draw,fill=white,minimum width=.6\columnwidth,below=0cm of attestorsign](attesteesign){\textit{Signature of Identity Owner}};
\draw[-latex] (metadata.north) -- (pchainentries.south);
\end{tikzpicture}
\caption{Claim Origin}
\label{subfig:claimorigin}
\end{subfigure}
\begin{subfigure}{0.45\textwidth}
\centering
\begin{tikzpicture}
\node[rectangle,draw,fill=lightgray!30!white,minimum width=.6\columnwidth](pchainentries){Previous Chain Entries};
\node[rectangle,draw,fill=white,minimum width=.6\columnwidth,below=.7cm of pchainentries](metadata){};
\node[rectangle,draw,fill=white,minimum width=.6\columnwidth,below=0cm of metadata](attestorsign){\textit{Signature of Attesting Party}};
\node[rectangle,draw,fill=white,minimum width=.6\columnwidth,below=0cm of attestorsign](attesteesign){\textit{Signature of Identity Owner}};
\draw[-latex] (metadata.north) -- (pchainentries.south);
\node [coordinate] (sink) [left=2 cm of metadata]{};
\draw[-latex] (metadata) to node [midway,above] {\scriptsize{Metadata Block}} (sink) ;
\end{tikzpicture}
\caption{Pure Attestation}
\label{subfig:pureattestation}
\end{subfigure}
\caption{Blockchain blocks in the Passive Model, without revocation}
\label{subfig:verifclaimmodel}
\end{figure*}
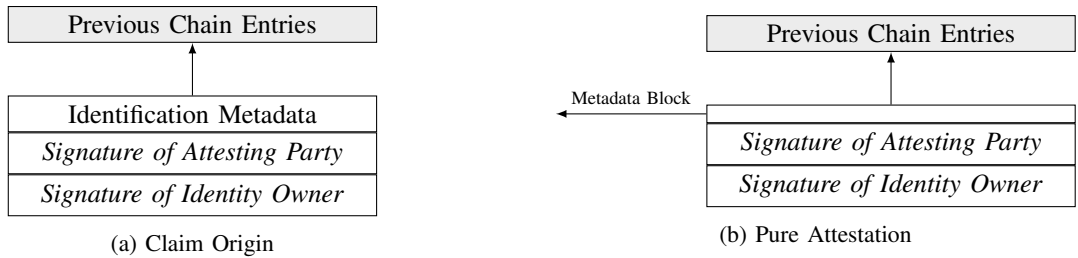
\begin{figure}[t]
\centering
\begin{tikzpicture}
\node[rectangle,draw,fill=lightgray!30!white,minimum width=.6\columnwidth](pchainentries){Previous Chain Entries};
\node[rectangle,draw,fill=white,minimum width=.6\columnwidth,below=.7cm of pchainentries](metadata){Identification Intent};
\node[rectangle,draw,fill=white,minimum width=.6\columnwidth,below=0cm of metadata](attestorsign){\textit{Signature of Verifier}};
\node[rectangle,draw,fill=white,minimum width=.6\columnwidth,below=0cm of attestorsign](attesteesign){\textit{Signature of Identity Owner}};
\draw[-latex] (metadata.north) -- (pchainentries.south);
\node [coordinate] (sink) [right=2 cm of metadata]{};
\draw[-latex] (metadata) to node [midway,above] {\scriptsize{Metadata Block}} (sink) ;
\end{tikzpicture}
\caption{Intent block for detectable revocation}
\label{subfig:verifintentmodel}
\end{figure}
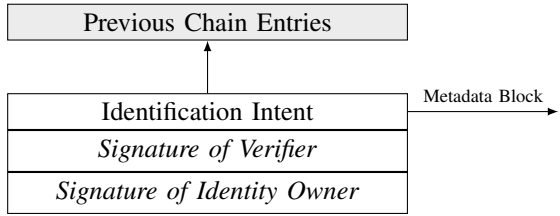
\begin{figure}[t]
\centering
\begin{tikzpicture}
\node[rectangle,draw,fill=lightgray!30!white,minimum width=.6\columnwidth](pchainentries){Previous Chain Entries};
\node[rectangle,draw,fill=white,minimum width=.6\columnwidth,below=.7cm of pchainentries](metadata){Intent Response};
\node[rectangle,draw,fill=white,minimum width=.6\columnwidth,below=0cm of metadata](attestorsign){\textit{Signature of Identity Owner}};
\node[rectangle,draw,fill=white,minimum width=.6\columnwidth,below=0cm of attestorsign](attesteesign){\textit{Signature of Attesting Party}};
\draw[-latex] (metadata.north) -- (pchainentries.south);
\node [coordinate] (sink) [left=2 cm of metadata]{};
\draw[-latex] (metadata) to node [midway,above] {\scriptsize{Intent Block}} (sink) ;
\end{tikzpicture}
\caption{Intent response block for real-time revocation}
\label{subfig:revocmodel}
\end{figure}
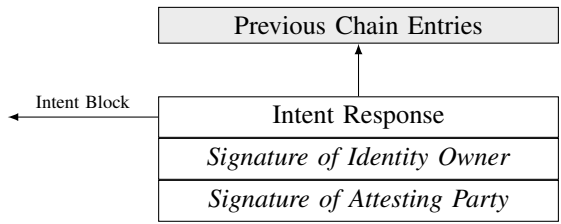

\subsubsection{Proof format}
Just like in OpenSSL certificates, we find the need to have a modular format to deal with advancements in the field of cryptography.
This means that if a better proofing method is developed, the system has minimal upgrading costs.
This is a reality in the field of Zero-Knowledge Proofs: in the span of just 2017 we have seen a several big advancements ranging from Ethereum's zk-SNARKs~\cite{mayer2016zk} to Stanford's Bulletproofs~\cite{bunz2017bulletproofs}.
Next to being able to upgrade easily, modular proofs also allow for different attributes to be proven through different Zero-Knowledge Proofs.
One might want a Zero-Knowledge Range proof for their ``age'' and a plain trapdoor function (note: which is not a Zero-Knowledge Proof) might be enough to store a first name.

\subsubsection{Proof link}
All of the properties of the metadata are voided if there is no link to the proof of the advertised attribute in the metadata.
Without a strong link to the proof, it would be possible to cheat.
This link must ensure that it is verifiable by third parties that the proof they play out with the identity owner is answered truthfully.
There are many ways to provide this verifiably truthful answering (one might for instance leverage MIT's Enigma~\cite{zyskind2015enigma} for publicly verifiable computation or use UCL's ClaimChain~\cite{kulynych2017claimchain} to tie this proof into the signature algorithm), but note that this mechanism will change as the method for the Zero-Knowledge Proof changes.

\subsection{Chains of claims}
Now that we have a definition for a claim which satisfies the \textbf{portability}, \textbf{interoperability}, \textbf{minimalization} and \textbf{protection} requirements, we will discuss how to integrate these into the blockchain structure.
Recall that we do this to provide the \textbf{transparency} and \textbf{persistence} needed for audit logs and the legal status of the identity system.

When designing this chain of claims, we recognize that there are different use cases for claims.
Some claims may be expected to last forever (once one hits the age of $21$, one will never become younger).
Some claims may not present an active risk when used whilst being revoked (receiving a parking permit for an address you no longer live at), even though they should be detected and punished.
Lastly, in the worst case, some claims may require real-time proof of correctness (not being a terrorist when checking in to the airport).
Depending on the use case for the attribute, we identify three levels of escalation for the audit logs: a \textit{passive}, an \textit{intent-based} and an \textit{active} model.
These approaches can be seen as escalations of each other as they each add an additional block to the chain, compared to the escalation level before them.
It should also be noted that these approaches are not mutually exclusive.
An identity chain can have each of these escalations in it, even for the same attribute.

\begin{table*}
\renewcommand{\arraystretch}{1.3}
\caption{Details of Self-Sovereign Identity implementations}
\label{tab:implementations}
\centering
\begin{tabular}{ l | l | l | l | l }
    \hline
    \textbf{Name} & \textbf{Context} & \textbf{Blockchain} & \textbf{Zero-Knowledge Proof} & \textbf{Language} \\  \hline
    IPv8 app & Dutch digital passport & TrustChain & Interactive Proof & Python \\
    IdentityChain & TU Delft course & TrustChain & Interactive (Range) Proof & Java, Kotlin \\
    Group A & TU Delft course & Proof-of-Work based & Non-Interactive Proof & Python \\
    Group B & TU Delft course & TrustChain & Non-Interactive Proof & Java \\
    \hline
\end{tabular}
\end{table*}

\subsubsection{Passive}
In \autoref{subfig:verifclaimmodel} a chain entry, or block, in the passive model is given.
These passive model blockchain blocks may come in two forms, either that of a \textit{claim origin} (\autoref{subfig:claimorigin}) or a \textit{pure attestation} (\autoref{subfig:pureattestation}).
As the name implies, the claim origin stores the original claim metadata (as previously discussed in \autoref{sec:design}) on the blockchain.
In the a pure model, this claim could be pushed to the blockchain without the existence of any attestation by an attesting party.
However, since no claim is worth anything without proof, as an optimization the block is only added to the chain when the first proof is collected.
Once the first attesting party signs for the claim, the \textit{double-signed} block is added to the chain.
For the same claim a user can then continue to receive attestations.
However, further attestations for the same claim need no longer store the entire claim.
Further attestations can simply sign for the block hash of the block containing the claim to avoid data duplication, as shown in \autoref{subfig:pureattestation}.

Recall that all of these blocks are under the full control of the identity owner.
This is why the last signature is always produced by the identity owner as he has the final say on whether or not to include this block in his own personalized blockchain.
This is due to the Self-Sovereign Identity requirements of \textbf{control} and \textbf{access}.

\subsubsection{Intent-based}
When considering applications which require an audit log for eventual fraud detection, the intent-based model can be used.
This builds on the passive model by adding an intent block to the blockchain structure, which is visualized in \autoref{subfig:verifintentmodel}.
As discussed in \autoref{sec:designgenclaims} for the timestamp field, this block can be used to make identity fraud detectable.
Any party wishing to audit the global logs, need but examine the revocation block and the intent block (both pointing to the same hash of a metadata block of the identity owner) to find the timestamp inconsistency.
Note that the verifying party can either store these blocks directly or wait for global consensus to be reached, before transacting with the identity owner.

\subsubsection{Active}
In the worst case, an active check for revocation is required.
This is not a good fit for the slow consensus protocols required for blockchain.
However, it is at the very least not worse than the existing method.
Regarding the liveness of the attesting party, instead of the offline transactions for the passive and intent-based models, we now require the attesting party to be online.
We can still leverage the intent-timestamp method to provide a unique challenge for the attesting party, to identify the identity owner.
By providing this unique challenge we show that the attesting party's response is current and not being replayed from an earlier response.
By this method the interaction is still just between (1) the verifying party and the identity owner and (2) the identity owner and the attesting party.
The attesting party's response is relayed by the identity owner and as such the identity owner has full \textbf{control} (he gives explicit \textbf{consent} for this message).
In other words, at no point do the attesting party and the verifying party communicate without the knowledge of the identity owner.

\section{Implementations}
\label{sec:implementations}
\autoref{sec:design} has presented a model which is blockchain agnostic and Zero-Knowledge Proof agnostic.
In support of this, an implementation is presented which uses a Proof-of-Work blockchain as well as implementations which use TrustChain.
Furthermore, these same implementations use different Zero-Knowledge Proof methods.
The different implementations have been summarized in \autoref{tab:implementations} and will be further detailed in the remainder of this section.

As some of the implementations do not have a name, we have assigned them a name (namely ``Group A'' and ``Group B'') for reference throughout the remainder of this paper.
Note that the IdentityChain, Group A and Group B implementations are based on student work for the Delft University of Technology 2017/2018 Blockchain Engineering course.
The implementations are available online through GitHub.

\begin{figure*}[t]
\centering
\includegraphics[width=\textwidth]{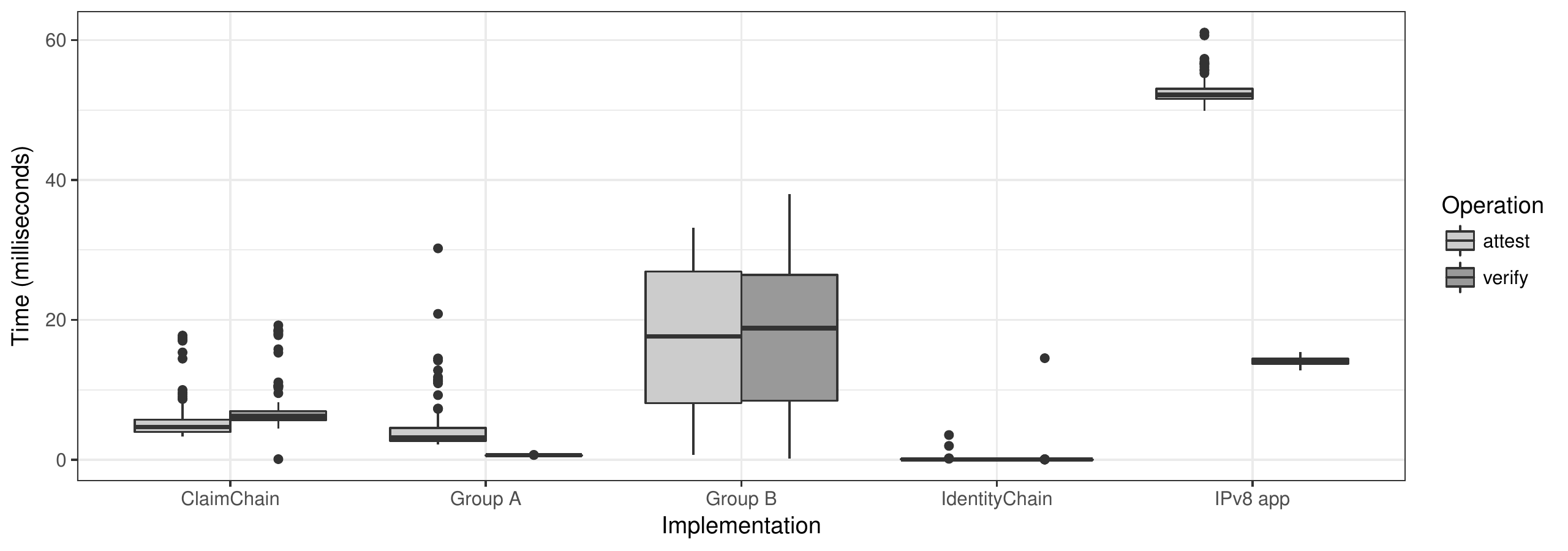}
\caption{Claim creation (left half: \textit{attest}) and claim verification (right half: \textit{verify}) times per implementation.}
\label{fig:impltimes}
\end{figure*}

The IPv8 Android app\footnote{https://github.com/Tribler/py-ipv8} (or \textit{IPv8 Attestation App} formally) leverages our own TrustChain structure as its personalized blockchain and uses a custom Zero-Knowledge Proofing structure.
The IPv8 app stems from the peer-to-peer networking framework called IPv8, which in itself is an amalgamation of functionality from the elastic database system Dispersy~\cite{zeilemaker2013dispersy} and the BitTorrent client Tribler~\cite{pouwelse2008tribler}.
The attestation system in the IPv8 app was the proof-of-concept which laid bare the concepts of this paper and was developed to serve as a solution for partial digitization of the Dutch passport.
It is currently being developed alongside claim-based biometry for passport-level authentication, made by the Dutch passport creator IDEMIA\footnote{https://www.idemia.com/}.
The details of this biometric system will remain proprietary.

The implementation uses unique keys per attribute stored in the blockchain.
The keys themselves are 1024-bit EC keys in $F_{p^2}$ without public message space $n$, but otherwise constructed as described by Boneh et al.~\cite{boneh2005evaluating} (over super-singular curve $y^2 = x^3 + 1$).
There is also another ed25519 key to publish data on the blockchain.
The exact implementation of the Zero-Knowledge Proofs will remain outside of the scope of this paper (the implementation is available online however).

The authors note that the key size is overkill and Python is not fit for either iOS or Android devices.
Therefore the IPv8 codebase is being ported to Java and will be made part of the TrustChain explorer app\footnote{https://play.google.com/store/apps/details?id=nl.tudelft.cs4160.trustchain\_android}.
The app will then also support different types of Zero-Knowledge Proofs, whereas the IPv8 app is currently limited to exact value proofs only.

IdentityChain is the first student work implemented in Java and Kotlin for Android.
The IdentityChain app\footnote{https://github.com/Tribler/tribler/issues/3236} requires a fingerprint to access the chain.
The supported Interactive Zero-Knowledge Proofs are single valued and range proofs, based on K. Peng and F. Bao's work~\cite{peng2010efficient}.
Only the default TrustChain ed25519 key is required for using this system.

The second student work, wich we dubbed Group A, implemented an identity system in Python with Javascript front end.
The app\footnote{https://github.com/Tribler/tribler/issues/3244} leverages a Bitcoin-like Proof-of-Work global blockchain for sharing identity instead of using a personalized blockchain structure.
Note that the \textbf{control} principle of Self-Sovereign Identities may waver if blocks are no longer added blindly to the chain.
A $\Sigma$-protocol implementation is then utilized to provide Non-Interactive Zero-Knowledge Proofs.

The final student work, Group B, implemented an identity system in Java for Android.
The app\footnote{https://github.com/Tribler/tribler/issues/3243}  shares attestations through QR-codes instead of using the Internet.
The supported Non-Interactive Proof is based on repeated hashing based on a shared secret value.
This system also only uses the TrustChain ed25519 key.

\section{Evaluation}
In this section the performance of the implementations mentioned in \autoref{sec:implementations} will be analyzed.
For real world deployment we expect both the provable claim creation as well as the claim verification to offer sub-second performance.
For each of the implementations in \autoref{tab:implementations} the creation time and the verification time has been measured.
The results of these measurements have been visualized as boxplots in \autoref{fig:impltimes}.
The boxplots have been grouped by implementation number, for which the time to create a claim (labeled as \textit{attest}, constituting the left boxplot of the pair) and the time to verify a claim (labeled as \textit{verify}, constituting the right boxplot of the pair) have been shown.
As a benchmark, the ClaimChain solution~\cite{kulynych2017claimchain} has also been measured and included in \autoref{fig:impltimes}.

For a fair comparison, the single value proof has been measured for each implementation.
The proof for a single set value is the only one supported by all of the implementations.
Furthermore, whereas the IPv8 app, Group A and Group B were capable of providing proofs for values of arbitrary length, the IdentityChain implementation was restricted to 20 bytes.
As such the input values for all of the implementations were sequences of 20 randomly sampled bytes.
Each of these inputs was then generated 100 times for each implementation to produce the visualized data.

Recall that the implementation languages for the IPv8 app, IdentityChain, Group A and Group B were Python, Kotlin, Python and Java respectively.
This has consequences for the timing measurements.
Measurements for both the Kotlin and Java programming languages were performed using the \texttt{System.nanoTime()} call, which provides a measurement granularity of nanoseconds.
Measurements within the Python programming language were performed using the \texttt{time.time()} call, which provides measurement granularity of $10^{-18}$ seconds.
As the machine running all of the experiments allocated one $3.20 GHz$ CPU core, the theoretical maximum precision is then of course only about $10^{-10}$.
Everything measured beyond the nanosecond scale in Python can therefore be assumed to be floating point error.
As all of the measured claim creation and verification times range in the dozens of milliseconds, we will claim this to be insignificant.

Moving on to the analysis of \autoref{fig:impltimes}, we immediately see that the claim creation time for the IPv8 app implementation is much higher than that of other solutions.
The most likely explanation for this, is that IPv8 creates a new key for each attribute.
As such this new key will also have to be sent with the claim, inflating the size of the claim and therefore also slowing down the verification process.
For the context of securing passport attributes, this might be a worthwhile performance trade-off though.

The IdentityChain implementation is by far the fastest of all the measured implementations, but also the most restrictive.
The implementation is not as widely usable as the other implementations, which do allow for arbitrary claim sizes.
This does show that a system tuned toward a specific input size is faster than a general solution.
In other words, when modeling data in the general provable claim format, the choice of a proof mechanism is vital for both claim creation speed and claim verification speed.

The Group B is the most interesting in terms of timing.
Whereas all of the other implementations have a significantly faster verification than creation time (even though it might not be visible, this also holds for the IdentityChain implementation), the verification and creation times do not differ significantly for the Group B implementation.
Also, the Group B implementation has the widest spread of claim verification times.
This makes the Group B implementation the most likely target for a side-channel attack, linking verification times to inputs.

Overall, disregarding the Group B implementation, we see that claim creation is significantly slower than claim verification.
This is a good property, as claim creation will only happen sporadically and claim verification will happen more often.
The real world analogue of this would be that one rarely receives a new passport, but one more frequently has to show it to others.
We also note that this does not hold for ClaimChain.

Lastly, it should be noted that all of the implementations achieve both claim creation as well as claim verification within 100 milliseconds.
This makes the technology easy to use for consumers and aids in the adoption of the technology, which was one of the targets of this paper.

\section{Related Work}
The concept of using blockchains as vessels for identity has been explained by Zyskind et al.~\cite{zyskind2015decentralizing}.
Generalizing over this concept, S. Azouvi et al.~\cite{azouvi2017secure} present a modern approach to Digital Identity through attestation.
This work builds upon that basic model of using digital signatures as attestations to the truth of data.
The added contribution of this paper over the S. Azouvi et al. paper is attribute value hiding and deployment in a permissionless and Self-Sovereign context.

In \textit{The Path to Self-Sovereign Identity}~\cite{allen2016principles} by C. Allen definitions of and requirements for User-Centric and Self-Sovereign Identity are given.
The essay supports the need for Self-Sovereign and User-Centric identity and explains why the current solutions do not meet the Self-Sovereign Identity requirements.
This paper fullfills all of the mentioned requirements of Self-Sovereign Identity.

In \textit{Blockchain-Enabled Self-Sovereign Identity}~\cite{wingerde2017thesis} by M.E.M. van Wingerde use-case driven requirements for blockchain based Self-Sovereign Identity are given.
These requirements have formed the basis of the Dutch government's stance and policy on Self-Sovereign Identity.
This paper can be seen as the accompanying realization of those requirements, forming a concrete model for Self-Sovereign Identity.

B. Kulynych et al. have presented \textit{ClaimChain}, an approach for chains of claims where the proof is part of the signature.
The paper presents its solution as a new standard, whereas this paper is more focused on providing an overarching abstraction.
The entire concept of \textit{ClaimChain} can then of course be modeled as an implementation of the model of this paper.
As such, the \textit{ClaimChain} would then also be interoperable with other methods for claim proof.
This also holds for other specialized systems like for example SCARAB as presented by Kokoris-Kogias et al.~\cite{kokorishidden}.

\section{Conclusion}
After 27 years, the web of trust finally has a chance to become reality.
This paper has presented the ground work for a Self-Sovereign Identity solution which is in production in the Netherlands for use by citizens in the latter half of 2018.
As such, this makes the solution of this paper the second (after Estonia) digitized passport solution to go live in the world.
The solution of this paper is the world's first \textit{permissionless decentralized} digitized passport and a true peer-to-peer identity commons.
For the first time, citizens will become the owners of their own identities.
No longer is the identity of users in the hands of a single (federated) authority.

Next to granting the users the properties of Self-Sovereign Identity, users are also prevented from cheating.
Verifying parties can efficiently capture an incriminating snapshot of the user's identity upon transacting.
Thus, if users decide to play out a proof which is incorrect, this will be detectable.
As users utilize legally valid signatures, this also means the individual behind the user's keys will be able to be punishable by law.

This paper has also presented methods of revocation of attestations to attribute value truth.
Through leveraging blockchain consensus rounds, revocations can be efficiently detected in the global network state.
This forms an efficient audit log which can be leveraged for pursuing legal action against users whom committed identity fraud.
For more severe applications, a real-time request of truth can also be needed.
This real-time solution does not improve upon current solutions, except for making sure that the flow of messages is always through the user.
Messages are never sent between parties the user transacts with, without knowledge and explicit consent of the user.

Implementations of the model of this paper are shown to have both sub-second claim creation and sub-second claim verification time on modern hardware.
This truly makes the Self-Sovereign Identity as portable as a passport, while retaining all of the benefits of digitization.
Thusly, we have created the new Dutch digital passport solution ready for global adoption.

\bibliographystyle{unsrt}
\bibliography{bibliography}{}

\end{document}